\documentclass[8pt,a4paper]{article}

\usepackage[utf8]{inputenc}     % Character encoding
\usepackage{amsmath, amssymb, amsfonts, amsthm} % Math
\usepackage{graphicx}           % Figures
\usepackage{subcaption}
\usepackage{booktabs, multirow} % Tables
\usepackage{float}              % Float placement
\usepackage{geometry}           % Page layout
\usepackage{setspace}           % Line spacing
\usepackage[numbers]{natbib}             % Bibliography
\usepackage{hyperref}           % Clickable references
\usepackage[table]{xcolor}      % Table colors
\usepackage{caption}            % Better captions
\usepackage[section]{placeins}  % Float control
\usepackage{authblk}
\usepackage{cleveref}
\usepackage{enumitem}
\usepackage{hhline}

\usepackage{tikz}
\usetikzlibrary{decorations.pathmorphing, patterns, shapes, decorations.pathreplacing, calligraphy}
\usepackage{pgfplots}
\pgfplotsset{compat=newest}

\numberwithin{equation}{section}

\theoremstyle{definition}

\graphicspath{ {./plots/} }

\title{Time-dependent two-way partial AUC and partial Youden Index estimator for right censored data}
\author[1]{Arpan Sanyal}
\author[2]{Sudheesh Kumar Kattumannil}
\author[1]{Ayon Ganguly \thanks{Corresponding author: \texttt{aganguly@iitg.ac.in}}}
\affil[1]{\small Department of Mathematics, Indian Institute of Technology Guwahati, Assam, India}
\affil[2]{\small Applied Statistics Unit, Indian Statistical Institute, Chennai, India}
\date{}

\begin{document}

\maketitle

\begin{abstract}
In medical research, it is often of interest to evaluate the predictive performance of a biomarker. Statistical approaches based on the Receiver Operating Characteristic (ROC) curve and its summary measures, such as the area under the curve (AUC) and the Youden index, are widely used to evaluate the prognostic performance of these biomarkers. In time-to-event studies, ROC analysis poses additional challenges due to change in disease status over time and the presence of censored individuals. To address these issues, time-dependent ROC curves were introduced. In this paper, we propose a non-parametric estimator of the time-dependent two-way partial AUC for right-censored data. We also discuss the partial Youden index and the associated optimal biomarker cutoff estimator for the right-censored data. We conduct an extensive simulation study to investigate the finite sample performance of the proposed estimators. The simulation study indicates that the proposed non-parametric estimators efficiently account for right censoring. Finally, we illustrate the proposed methods using two real data sets, one from the Primary Biliary Cirrhosis study and the other from the Molecular Taxonomy of Breast Cancer International Consortium trial.
\end{abstract} 


\section{Introduction}

In modern clinical studies, biomarkers (or diagnostic/prognostic tests/risk scores) are often used to identify individuals who are more likely to experience a clinical event of interest, such as disease onset, recurrence, recovery, or death. For example, the ``Framingham risk score" is a well-known biomarker used to identify patients at risk of developing coronary heart disease in $10$ years.\cite{frambiom} The biomarker is obtained using risk factors such as age, blood pressure, smoking status, cholesterol level, diabetes, high density lipoproteins. Identifying these individuals is crucial for early prevention, treatment planning, and drug development. However, before using a newly identified biomarker in clinical practice, it is necessary to evaluate its diagnostic performance and compare it with that of existing biomarkers. Receiver Operating Characteristic (ROC) curve analysis is a widely used statistical technique for assessing the discriminatory ability of these biomarkers.\cite{pepe2003statistical, Zhou2011} A ROC curve is a graphical plot of true positive rate (TPR) and the corresponding false positive rate (FPR) for all the possible threshold values of the biomarker. The TPR represents the probability of correctly classifying individuals who truly experience the event of interest, while the FPR represents the probability of incorrectly classifying individuals who do not experience the event of interest. The area under the ROC curve (AUC) is a popular summary metric for ROC analysis. It provides a global summary of biomarkers' performance in distinguishing between classes in a binary classification study across the entire ROC space. It is equivalent to the probability that a randomly selected individual with the event of interest is ranked higher than a randomly selected individual without the event of interest.\cite{Dodd01062003} Given two biomarkers, the biomarker with a larger AUC has better overall predictive performance. In practice, when diagnosing a lethal disease such as cancer studies, it is crucial to achieve a high TPR to correctly identify severely diseased subjects. Moreover, an elevated FPR will result in the unnecessary use of scarce medical resources on healthy individuals. Hence, from a clinical perspective, the region of the ROC curve with low FPR and high TPR is of primary interest. Most of the existing literature suggests restricting the FPR by an upper bound and a lower bound to consider a partial area under the ROC. This is referred to as FPR pAUC in the literature.\cite{doddpepebiom,useofpaucstatmed} The upper bound of FPR is to maintain the FPR values below a maximum tolerable threshold, and the lower bound of FPR is used to indirectly maintain the minimum desired TPR in a diagnostic test.

\begin{figure}[ht]
    \centering 
    \begin{minipage}[b]{0.48\textwidth}
        \includegraphics[width=\textwidth]{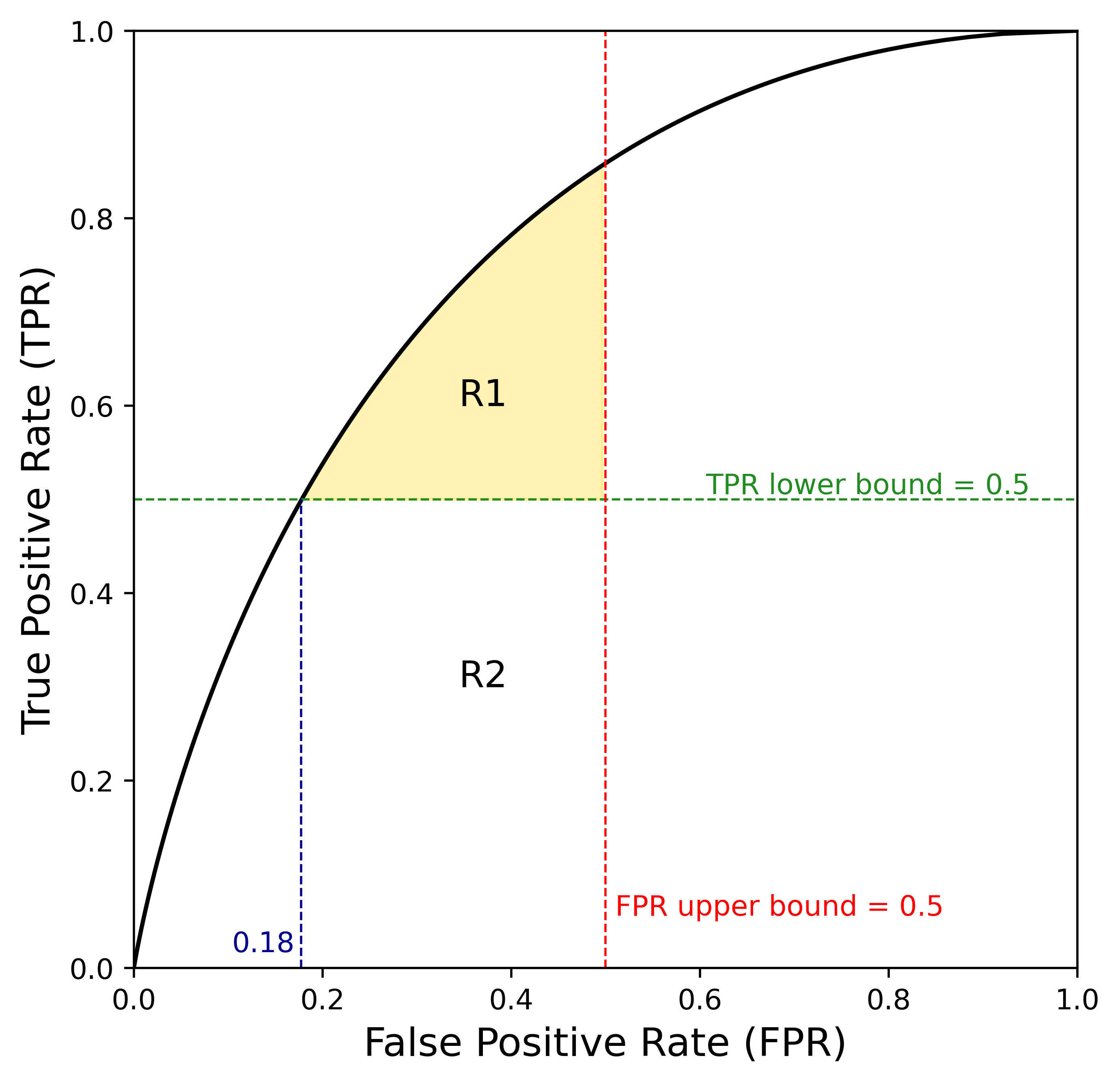}
    \end{minipage}
    \begin{minipage}[b]{0.48\textwidth}
         \includegraphics[width=\textwidth]{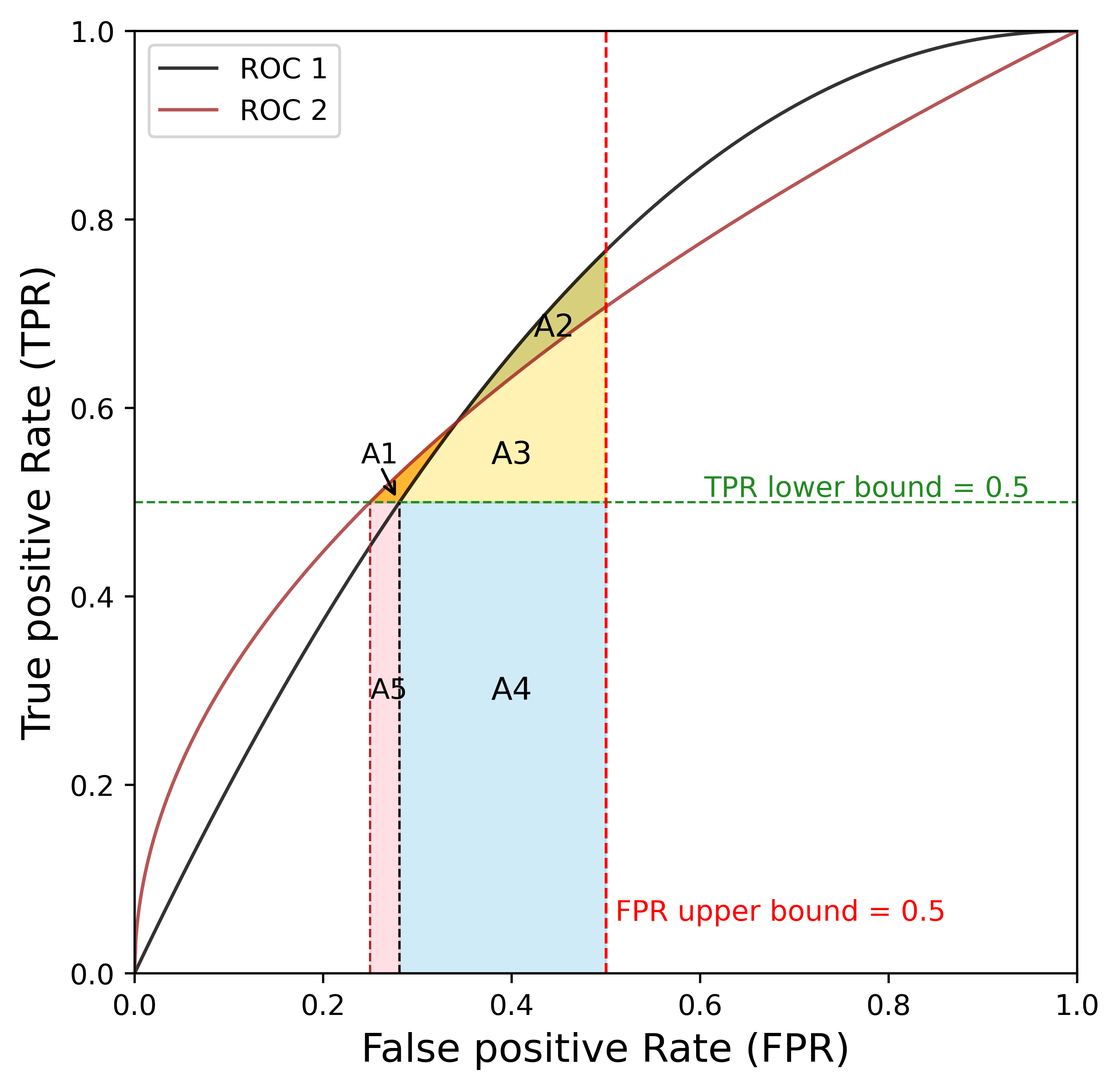}
    \end{minipage}      
   \caption{A visual illustration of two-way partial AUC.}
    \label{fig:combined}
\end{figure}

Yang et al\cite{twowaypaucstatmed} demonstrated such indirect control over TPR results in considering the redundant area below the desired TPR level. To overcome these issues in FPR pAUC, they introduced two-way pAUC (TPAUC), which directly controls both TPR and FPR. In TPAUC, we directly bound the TPR by a lower bound to maintain the minimum desired TPR, and the FPR by an upper bound to keep FPR below a maximum tolerable threshold. For example, consider the ROC curves in Figure~\ref{fig:combined}. The shaded region (R1) in Figure 1(a) is the area corresponding to the two-way pAUC defined by a TPR lower bound of 0.5 and FPR upper bound of 0.5, respectively. In contrast, the area corresponding to the FPR pAUC is $\mathrm{R1 +R2}$, as it indirectly maintains the same TPR lower bound of 0.5, by setting the FPR lower bound of 0.18 (blue dotted line). Hence, it also incorporates a redundant area R2 below the desired TPR level. This is conceptually inconsistent because it violates the original intention to exclude low-TPR regions. Considering this redundant area may also lead to inaccurate conclusions. For example, consider the scenario in Figure 1(b), which contains two intersecting ROC curves, denoted as ROC 1 (black) and ROC 2 (red), corresponding to two distinct biomarkers. The region of interest is where TPR $\geq 0.5$ and FPR $\leq 0.5$. As $\mathrm{A2 > A1}$, the ROC 1 demonstrates better diagnostic ability than ROC 2 within this region of interest. For the two-way pAUC, the areas corresponding to ROC 2 and ROC 1 are $\mathrm{A2 + A3}$ and $\mathrm{A1 + A3}$, respectively. Since $\mathrm{A2 + A3 > A1 + A3}$, the two-way pAUC correctly identifies ROC 1 as having better performance in the region of interest. In contrast for the FPR-based pAUC, the areas under ROC 1 and ROC 2 are $\mathrm{A2 + A3 + A4}$ and $\mathrm{A1 + A3 + A4 + A5}$, respectively. Because $\mathrm{A5 + A2 > A1}$, the FPR-based pAUC incorrectly suggests that ROC 2 outperforms ROC 1, leading to a misleading interpretation of diagnostic performance. This misinterpretation arises as the FPR-based pAUC includes the redundant area $\mathrm{A4}$ and $\mathrm{A5}$ in its calculations. This suggests that two-way pAUC is more efficient and accurate than FPR pAUC. Similar to FPR pAUC, another variant of pAUC, called TPR pAUC, exists in the literature.\cite{doddpepebiom} In TPR pAUC, the TPR is restricted by upper and lower bounds, where the upper bound of TPR is used to indirectly maintain FPR below a tolerable threshold. Therefore, TPR pAUC also suffers from the same misleading interpretation as FPR pAUC.\cite{twowaypaucstatmed} In what follows, we shall refer to TPR pAUC and FPR pAUC as classical pAUC. For a detailed theoretical demonstration and simulations comparing classical pAUC and two-way pAUC can be found in their work.

In addition to AUC, another popular summary measure for ROC analysis is the Youden index. It is a linear function of TPR and FPR, which provides a direct measure of the highest diagnostic accuracy achievable by a biomarker. The biomarker value corresponding to the Youden index yields the optimal balance between the highest possible TPR and the lowest possible FPR, and this biomarker value is referred to as the optimal biomarker cutoff point. Classically, the Youden index is defined over the entire ROC space. Motivated by the clinical advantages of pAUC, Li et al. \cite{partialyouden} introduced a new summary metric of the ROC curve, called the partial Youden index. It is used to assess a biomarker's highest diagnostic accuracy in a region of clinical interest, i.e. the region of the ROC curve with high TPR and low FPR. Li et al. \cite{partialyouden} emphasized its importance in precision medicine, highlighting that, analogous to the pAUC, the partial Youden index provides a more clinically meaningful assessment of biomarker's diagnostic performance when attention is restricted to specific regions of clinical interest of the ROC curve.

Unlike in classical ROC analysis, where the event status is known in advance, in prognostic studies such as time-to-event studies, the event status of the subjects varies with time. In addition to time-dependent events, some subjects in time-to-event studies may be censored during the study. Hence, it is necessary to incorporate a time dimension and censoring information in ROC analysis for time-to-event studies.  To address different clinical problems, Heagerty et al.  \cite{heagertybiometrics} and Heagerty and Zheng\cite{Heagertyzheng2005} proposed, in their seminal works, three time-dependent extensions of the classical ROC, viz., cumulative/dynamic, incident/dynamic, and incident/static time-dependent ROC. These approaches are constructed by dichotomising time-dependent events into case and control, using three different risk-set definitions at a given time point $t$. Over the past few decades, several authors, including  Heagerty et al.  \cite{heagertybiometrics}, Heagerty and Zheng\cite{Heagertyzheng2005},  Chambless and Diao\cite{chambless2006estimation},  Uno et al.\cite{uno2007evaluating}, Chiang
and Hung\cite{CHIANG20101162}, Lambert et al.\cite{lambertcd} Mart\'{\i}nez-Camblor et al.\cite{Martínez-Camblor21112016}, Li et al.\cite{Lietal},  Beyene and El Ghouch\cite{beyene2020smoothed}, Wu et al.\cite{wu2020predictive}, Nu\~{n}o and Gillen\cite{censorrobust2021} have proposed different estimators for time-dependent TPR, FPR, and AUC under various censoring scenarios for these three extensions of time-dependent ROC. A comprehensive review of the works in last decades can be found in the review papers of Blanche et al.\cite{blanche2013review} and Kamarudin et al.\cite{kamarudin2017time}.

 Although an ample amount of literature exists on the estimation and analysis of the time-dependent ROC curve, comparatively little attention has been devoted to the partial region of interest. Despite the clinical advantages of the pAUC, the time-dependent pAUC remains underexplored. In cumulative/dynamic time-dependent extensions of ROC, Chiang and Hung\cite{CHIANG20101162} proposed an Akritas-based AUC estimator for the time-dependent ROC curve. Later, Hung
and Chiang\cite{hung2011nonparametric} extended this Akritas-based AUC estimator for time-dependent FPR-pAUC when the FPR is bounded by an upper bound. Recently, Jiang et al. \cite{jiang2024analyzing} proposed an FPR-pAUC estimator for the incident/dynamic extension of time-dependent ROC. To the best of our knowledge, these two are the only existing works that address FPR-pAUC for two distinct extensions of time-dependent ROC. However, as these works extend the classical FPR-based pAUC, therefore they inherit the limitations of FPR-pAUC. 
 
Owing to the advantages over the classical pAUC and Youden index, the TPAUC and partial Youden index have recently attracted the attention of several researchers in the field of classical ROC analysis, including Yang et al. \cite{yang2022optimizing} Lavazza and Morasca\cite{lavazza2022considerations}, Wechsung and Konietschke \cite{wechsung2023simultaneous}, Chaibub Neto et al. \cite{chaibub2024novel}, Jia et al.\cite{partialyoudenverificationbias}. However, in spite of the popularity in classical ROC analysis, the TPAUC and partial Youden index have not been explored for time-dependent ROC. To fill this research gap, in this paper, we propose time-dependent partial area under the curve and partial Youden index estimators for cumulative/dynamic ROC for right-censored data. To the best of our knowledge, this is the first extension of TPAUC and partial Youden index for time-dependent ROC. The proposed estimators are completely non-parametric in nature and efficiently account for right censoring. Moreover, they inherit the advantages of their classical counterparts, offering a more efficient and accurate evaluation of predictive performance within a region of interest in the ROC space compared to the existing methods. 

 The rest of the manuscript is organised as follows. In Section 2, we discuss the definitions and estimation methods for time-dependent TPR, FPR, $\mathrm{TPAUC}$, the partial Youden index, and the corresponding optimum cutoff point. We then discuss the inference procedure for the proposed estimators in the same section. In Section 3, we present an extensive simulation study to evaluate the finite-sample performance of the proposed estimator and the bootstrap-based inference procedure. In Section 4, we illustrate our proposed method on two real-life datasets. Finally, in Section 5, we conclude the article with a discussion and directions for future research.    

\section{Methods}
\label{sec:methods}
Let $T$ and $M$, respectively, denote the time to event of interest and a continuous biomarker of an individual or a subject. For a given time horizon $t$, $D_t$ = $I(T\leq t)$ denotes the event status of the subject, where 
     \begin{align*}
        I(A) = \begin{cases}
           1 & \text{if } A \text{ is true}\\
           0 & \text{if } A \text{ is false}.
        \end{cases}
     \end{align*}
     Therefore, $D_t$ takes the value 1 or 0, indicating the presence or absence of the event of interest on or before time $t$. Subjects with $D_t$ = 1 are referred to as cases, and those with $D_t$ = 0 are referred to as controls. We assume that a higher value of biomarker $M$ is associated with a higher risk of experiencing the event of interest. For a possible biomarker threshold value $m\in \mathbb{R}$, the time-dependent TPR and FPR for cumulative/dynamic ROC at time horizon $t > 0$ is given by Heagerty et al. \cite{heagertybiometrics}

\begin{align}
    \mathrm{TPR_t}(m) &= P(M > m \mid T \leq t), \nonumber \\
    \mathrm{FPR_t}(m) &= P(M > m \mid T > t).
    \label{1}
\end{align}

\noindent The time-dependent ROC curve , $\mathrm{ROC}_t$ , at time horizon $t > 0$, is obtained by plotting $\mathrm{TPR_t}$  and the the corresponding $\mathrm{FPR_t}$  for all possible biomarker value $m$. Mathematically, it can be written as
\begin{align}
\mathrm{ROC}_t(p) 
= \mathrm{TPR}_t \bigl\{ \mathrm{FPR}_t^{-1}(p )\}, 
\quad p \in [0,1],
\end{align}
where ${\mathrm{FPR}}_t^{-1}(p) = \inf\{ m \mid{\mathrm{FPR}}_{t}(m) \leq p\}$,

For a given TPR lower bound $l_0$ and FPR upper bound $u_0$, we define the time-dependent $\mathrm{TPAUC}$ at time horizon $t > 0$ as follows: 
\begin{align}
\mathrm{TPAUC}_t(l_0,u_0) 
&= P\!\left(
M_{1} > M_{2}, \;
M_{1} \leq \mathrm{TPR}_t^{-1}(l_0) , M_{2} \geq \mathrm{FPR}_t^{-1}(u_0)
\;\middle|\;
{T}_{1} \leq t, \; {T}_{2} > t\right),
\label{tpa}
\end{align}
where $M_1$, $M_2$ and $T_1$, $T_2$ represent the biomarker values and the event times for two different randomly selected individuals, respectively.\\
The time-dependent partial Youden index is defined by
\begin{align}
    \mathrm{PJ_t}(l_0,u_0) = \max_{m}\{\mathrm{TPR_t}(m) - \mathrm{FPR_t}(m) \mid \mathrm{TPR_t}(m) \geq l_0 , \mathrm{FPR_t}(m) \leq u_0 \}.
    \label{pj}
\end{align}

\noindent The biomarker value corresponding to the partial Youden index $\mathrm{PJ_t}(l_0,u_0)$ is denoted by $m^{\mathrm{opt}}_t(l_0,u_0)$, and is referred to as the optimal cutoff value. It is defined as,

\begin{align}
    m^{\mathrm{opt}}_t(l_0,u_0) = \underset{m}{\operatorname{arg\max}}\{\mathrm{TPR_t}(m) - \mathrm{FPR_t}(m) \mid \mathrm{TPR_t}(m) \geq l_0 , \mathrm{FPR_t}(m) \leq u_0 \}.
    \label{pj_opt}
\end{align}

\subsection{Estimation}
Assume that the time-to-event $T$ and the censoring time $C$ are conditionally independent given the biomarker $M$. We observe $n$ independent samples of the form $\{O_i=(Y_i, \Delta_i, M_i), \; i = 1, 2, \ldots, n\}$ , where $Y_i = \min(T_i,C_i)$  is the observed time for the $i$th subject , $ \Delta_i = I(T_i\leq C_i)$ is the censoring indicator for the $i$th subject. In the classical, non-censored ROC setting, the TPR and FPR can be directly estimated since the event status $D_t$ is known for all subjects. However, when the survival time $T$ is subject to right censoring, the event status $D_t$ remains unknown for some subjects.  Mart\'{\i}nez-Camblor et al.\cite{Martínez-Camblor21112016} and Li et al.\cite{Lietal} independently proposed estimating the event status of a right-censored subject by introducing weights. The weight at time $t>0$ of a subject is the probability of being a non-survivor at time $t$ given the observed data. Thus, the weight at time $t>0$ can be written as,
\begin{equation}
W_{1}(t) = P(T \leq t \mid M, \Delta, Y )  
= \left[ 1 - (1 - \Delta) \frac{S(t \mid M)}{S(Y \mid M)} \right] 
I(Y \leq t),
\label{weight}
\end{equation}
where $S(.|M)$ is the conditional survival function of $T$ given the biomarker $M$.\\
\noindent Let 
     \begin{align*}
        W_0(t) = 1-W_1(t) = P \left( T>t \mid M, \Delta, Y \right).
     \end{align*}  The conditional survival function of T given M can be estimated from the observed data using the Beran estimator.\cite{beran1981nonparametric} The Beran estimator  is defined as,

\[
\widehat{S}(t|m) = \prod_{i: Y_i \leq t} \left( 1 - 
\frac{\psi_i(m)}{\sum_{j=1}^n \psi_j(m) I(Y_j \geq Y_i)} \right)^{\Delta_i}, 
\]
where $\psi_i(m) = \frac{ k\!\left( \frac{M_i - m}{b} \right) }
{ \sum_j k\!\left( \frac{M_j-m}{b} \right) },$ $b$ is bandwidth and $k(.)$ is a kernel function. Plugging in the estimated conditional survival functions $\widehat{S}(.|m)$ in \eqref{weight} we obtain the weight estimates as $\widehat{W}_1(t)$ and $\widehat{W}_0(t)$, respectively. For our method, we have used a Gaussian kernel, defined by,
\begin{align}
K(u) &= \frac{1}{\sqrt{2\pi}} e^{-\frac{1}{2}u^2}
\end{align}
For bandwidth selection, we have used data-driven bandwidth selection based on a plug-in method.\cite{sjband}

\noindent Using these estimated weights, we can estimate $\mathrm{TPR_t}(m)$ and $\mathrm{FPR_t}(m)$ as follows
\begin{align}
\widehat{\mathrm{TPR}}_{t}(m) 
= \frac{1 }{\widehat{N}_1(t)}\sum_{i=1}^n I(M_i > m)\widehat{W}_{i1}(t),
\label{etp}
\end{align}
and
\begin{align}
\widehat{\mathrm{FPR}}_{t}(m) 
= \frac{1}{\widehat{N}_0(t)}\sum_{i=1}^n  I(M_i > m)\widehat{W}_{i0}(t),
\label{efp}
\end{align}
where $\widehat{N}_1(t) = \sum_{i=1}^n \widehat{W}_{i1}(t)$, $\widehat{N}_0(t) = \sum_{i=1}^n \widehat{W}_{i0}(t) = n-\widehat{N}_1(t)$.

 Let $\widehat m_p=\widehat{\mathrm{TPR}}_t^{-1}(l_0) = \sup\{ m \mid \widehat{\mathrm{TPR}}_{t}(m) \geq l_0\} $ and $\widehat m_w=\widehat{\mathrm{FPR}}_t^{-1}(u_0) = \inf\{ m \mid \widehat{\mathrm{FPR}}_{t}(m) \leq u_0\} $. Then from \eqref{tpa}, $\mathrm{TPAUC}_t(l_0,u_0)$ can be estimated by
\begin{align}
\widehat{\mathrm{TPAUC_t}}(l_0,u_0) 
= \frac{ \sum_{i=1}^n \sum_{j=1}^n \widehat{W}_{i1}(t)\widehat{W}_{j0}(t) 
\left[ I(M_i > M_j) + \tfrac{1}{2} I(M_i = M_j) \right] 
I(M_i \leq \widehat m_p) I(M_j \geq \widehat m_w) }
{ \widehat{N}_0(t)\widehat{N}_1(t) }.
\label{etpauc}
\end{align}

 \noindent Note that the terms $\frac{1}{2} I \left( M_{i}=M_j \right)$ are introduced in \eqref{etpauc} to incorporate ties in biomarkers.
The time-dependent partial Youden index can be estimated as
\begin{align}
    \widehat{\mathrm{PJ_t}}(l_0,u_0) &= \max_{m}\{\widehat{\mathrm{TPR}}_{t}(m) - \widehat{\mathrm{FPR}}_{t}(m) \mid \widehat{\mathrm{TPR}}_{t}(m) \geq l_0 , \widehat{\mathrm{FPR}}_{t}(m) \leq u_0 \}  \nonumber\\
    &=\max_{\widehat m_p\leq m\leq \widehat m_w} \left\{\frac{1 }{\widehat{N}_1(t)}\sum_{i=1}^n I(M_i > m)\widehat{W}_{i1}(t) - \frac{1}{\widehat{N}_0(t)}\sum_{i=1}^n  I(M_i >m)\widehat{W}_{i0}(t)\right\}.
    \label{epj}
\end{align}
The estimator of corresponding optimal biomarker cutoff is
     obtained as
     \begin{align*}
        \widehat{m_t^{\text{opt}}} (l_0,\,u_0) = \underset{\widehat m_p\leq m\leq \widehat m_w}{\operatorname{arg\,max}}
        \left\{\frac{1 }{\widehat{N}_1(t)}\sum_{i=1}^n I(M_i > m)
        \widehat{W}_{i1}(t) - \frac{1}{\widehat{N}_0(t)}\sum_{i=1}^n  I(M_i >m)
        \widehat{W}_{i0}(t)\right\}.
     \end{align*}

 A theoretical derivation of estimators \eqref{etp}, \eqref{efp}, \eqref{etpauc} and \eqref{epj} can be found in the Appendix \ref{app:dervation}. For notational simplicity, in what follows, we suppress the parameters $l_0$ and $u_0$ from the notation of the estimators.

\subsection{Inference }

\subsubsection{Confidence Intervals}

We adopt a nonparametric bootstrap procedures to obtain inference for our proposed estimators. \\

Let $\hat{\theta}_t^{(b)}$ denote the estimate of $\theta_t$ obtained from the $b$-th bootstrap sample, $b = 1, \ldots, B$. The parameter $\theta_t$ may represent any of the estimands $\mathrm{TPAUC}_{t}$, $\mathrm{PJ_t}$, or  $m^{\mathrm{opt}}_t$. A $100(1-\alpha)\%$ percentile bootstrap confidence interval for $\theta_t$ is obtained as
\[
\left[ \, \hat{\theta}_t^{(\alpha/2)}, \; \hat{\theta}_t^{(1-\alpha/2)} \, \right],
\]
where $\hat{\theta}_t^{(\alpha/2)}$ and $\hat{\theta}_t^{(1-\alpha/2)}$ are the $(\alpha/2)$th and $(1-\alpha/2)$th empirical quantiles of the bootstrap distribution $\{\hat{\theta}_t^{(1)}, \ldots, \hat{\theta}_t^{(B)}\}$. For brevity, in what follows we shall only report bootstrap estimates for $\mathrm{TPAUC}_{t}$.

\section{Simulation Study}
\subsection{Simulation Setup}
We generate the survival time $T$ from a Weibull proportional hazards model with 
scale and shape parameters fixed at $1.5$ and $2$, respectively. The conditional hazard function is given by 

\[
h(t \mid X)
= \frac{2}{1.5}\left(\frac{t}{1.5}\right)\exp\!\left(\beta X\right),
\qquad t > 0,
\]

\noindent where, $X \sim \mathcal{N}(0,1)$ 
represents the covariate, and $\beta$ is the regression coefficient associated with $X$. hence, the survival time is obtained using the cumulative  hazard inversion  given by
\[
    T \;=\; 1.5 \left[-\log(U)\exp\!\left(-\beta X\right)\right]^{1/2},
\]
\noindent where $U \sim \mathrm{Uniform}(0,1) $. 
 
The biomarker is defined as
\[
    M \;=\; \exp\!\left\{-\left(\tfrac{T}{1.5}\right)^{2} \exp\!\left(\beta X\right)\right\}.
\]
Note that this is the survival function of the Weibull proportional hazards model.
We consider two values of the regression coefficient, $\beta = -0.5$ and $\beta = -0.75$, 
where the higher value corresponds to a stronger association between $T$ and $M$.\\
\noindent
Finally, the censoring time \(C\) is generated from an exponential distribution with the probability density function
\[
f_C(c)= \dfrac{1}{\lambda_C}\exp\!\left(-\dfrac{c}{\lambda_C}\right) \quad c> 0, 
\]
 We consider two values of the scale parameter, \(\lambda_C = 2.9\) and \(\lambda_C = 1.8\), to achieve censoring rates of \(35\%\) and \(50\%\), respectively. 
 % A similar setup has also been considered by  Dey et al.\cite{Dey2023InferenceMeasures},  Beyene et al.\cite{kmbprc}.

To assess the performance, we generate $N = 1000$ datasets of sample sizes $n = 500$ and $n = 1000$. We consider three time horizons $t$ . For each scenario, we report the AE, standard deviation (SD), absolute bias (Bias), MSE, average bootstrap standard deviation (ASD), AW and CP of $95\%$ percentile bootstrap confidence intervals from 1000 bootstrap samples. The true values of $\mathrm{TPAUC}_t$,  $\mathrm{PJ_t}$, and $m^{\mathrm{opt}}_t$ are obtained using Monte Carlo method based on a large uncensored simulated dataset of size $100000$, generated under the aforementioned setup. As discussed in the introduction, in clinical applications, a biomarker with TPR greater than 50\% and FPR less than 50\% is desired. To assess the performance of the proposed estimators and bootstrap inference procedure in such a region of clinical interest in ROC space, we consider a $\mathrm{TPR}$ lower bound of $l_0=0.50$  and $\mathrm{FPR}$ upper bound $u_0=0.40$ .

\subsection{Simulation Results}

\begin{table}[h]
\centering
\caption{Simulation results for the $\mathrm{TPAUC}_t$ estimator: Estimates, Bias ($\times$100), MSE ($\times$1000), SD, bootstrap ASD, AW and CP of $95\%$ bootstrap confidence intervals for different sample sizes ($n$), censoring rates (cens), and prediction times ($t$).}
\vspace{2mm}
\small
\label{T1}
\setlength{\tabcolsep}{5pt} % column spacing
\renewcommand{\arraystretch}{2}
\begin{tabular}{ccc|cccccccccc}
\toprule
$n$ & cens & $t$ & True & Estimate & SD & Bias$(\times 10^2)$ & MSE$(\times 10^3)$ & ASD & AW & CP \\
\midrule
\multicolumn{11}{l}{$\beta = -0.5$} \\
\multirow{6}{*}{500} 
& 35 & 0.9 & 0.1637 & 0.1613 & 0.0079 & 0.2362 & 0.0683 & 0.0082 & 0.032 & 0.957 \\
& 35 & 1.2  & 0.1529 & 0.1512 & 0.0091 & 0.1738 & 0.0856 & 0.0091 & 0.035 & 0.948 \\
& 35 & 1.6 & 0.1443 & 0.1439 & 0.0101 & 0.0371 & 0.1015 & 0.0105 & 0.041 & 0.951 \\
& 50 & 0.9 & 0.1637 & 0.1603 & 0.0087 & 0.3382 & 0.0871 & 0.0087 & 0.034 & 0.945 \\
& 50 & 1.2   & 0.1529 & 0.1504 & 0.0101 & 0.2538 & 0.1074& 0.0099 & 0.038 & 0.935\\
& 50 & 1.6 & 0.1443 & 0.1431 & 0.0116 & 0.1191 & 0.1367 & 0.0118 & 0.046 & 0.943 \\
\cmidrule(lr){2-11}
\multirow{6}{*}{1000} 
& 35 & 0.9 & 0.1637 & 0.1624 & 0.0057 & 0.1264 & 0.0339 & 0
0057& 0.022 & 0.954 \\
& 35 & 1.2 & 0.1529 & 0.1517 & 0.0064 & 0.1238 & 0.0419 & 0.0063 & 0.025 &  0.944\\
& 35 & 1.6 & 0.1443 & 0.1447 & 0.0072 &0.0409  & 0.0513 & 0.0073 & 0.028 & 0.94 \\
& 50 & 0.9 & 0.1637 & 0.1620 & 0.0059 & 0.1670 & 0.0379 & 0.0060 & 0.024 & 0.945 \\
& 50 & 1.2 & 0.1529 & 0.1513 & 0.0068 & 0.1625 & 0.0489 &0.0069  & 0.027 & 0.949 \\
& 50 & 1.6 & 0.1443 & 0.1441 & 0.0081 & 0.0205 & 0.0656 &0.0082  & 0.032 & 0.95 \\
\midrule
\multicolumn{11}{l}{$\beta = -0.75$} \\
\multirow{6}{*}{500} 
& 35 & 0.9 & 0.1287 & 0.1280 & 0.0111 & 0.0722 & 0.1242 & 0.0111 & 0.043 & 0.948 \\
& 35 & 1.2   & 0.1156 & 0.1154 & 0.0118 & 0.0249 & 0.1399 & 0.0116 & 0.045 & 0.934 \\
& 35 & 1.6 & 0.1088 & 0.1090 & 0.0134 & 0.0226 & 0.1798 & 0.0128 & 0.050 & 0.936 \\
& 50 & 0.9 & 0.1287 & 0.1272 & 0.0121 & 0.1523 & 0.1492 & 0.0118 & 0.046   & 0.942 \\
& 50 & 1.2   & 0.1156 & 0.1145 & 0.0128 & 0.1087 & 0.1658 & 0.0126 & 0.049 & 0.947 \\
& 50 & 1.6 & 0.1088 &   0.1084 & 0.0150 & 0.0442 & 0.2251& 0.0144 & 0.056 & 0.934 \\
\cmidrule(lr){2-11}
\multirow{6}{*}{1000} 
& 35 & 0.9 & 0.1287 &0.1282  & 0.0077 & 0.0534 & 0.0600 & 0.0079 & 0.031 & 0.945  \\
& 35 & 1.2 & 0.1156 & 0.1159 & 0.0085 & 0.0327 & 0.0719 & 0.0082 & 0.032 & 0.936 \\
& 35 & 1.6 & 0.1088 & 0.1097 & 0.0091 & 0.0911 & 0.0833 & 0.0090 & 0.035 & 0.945 \\
& 50 & 0.9 & 0.1287 & 0.1277 & 0.0086 & 0.0932 & 0.0750 & 0.0084 & 0.033 & 0.938  \\
& 50 & 1.2 & 0.1156 & 0.1155 & 0.0092 & 0.0053 & 0.0848 &0.0089  & 0.035 & 0.939 \\
& 50 & 1.6 & 0.1088 & 0.1091 & 0.0105 & 0.0335 & 0.1095 & 0.0101 & 0.040 & 0.946 \\
\bottomrule
\end{tabular}
\end{table}

%%%%%%%%%%%%%%%%%%%%%%    simulation tables end    %%%%%%%%%%%%%%%%%%%%%%

Here, we present the simulation results in Tables~\ref{T1}--\ref{T2}, based on the data generation scheme described above. 
Tables~\ref{T1} report the results for $\widehat{\mathrm{TPAUC}}_{t}$, 
Tables~\ref{T2} present the results for $\widehat{\mathrm{PJ}}_{t}$, 
and the estimated optimal cutoff values. The tables are based on different sample sizes ($n$), censoring proportions (cens), and prediction time horizons ($t$).

Table~\ref{T1}, present the simulation results for $\widehat{\mathrm{TPAUC}}_{t}$ . From the tables we observe that for  $\widehat{\mathrm{TPAUC}}_{t}$ estimator the bias was less than $0.3382(\times 10^{-2})$ and the MSE was less than $0.2251 (\times10^{-3})$.From ~\ref{T2} for $\widehat{\mathrm{PJ}}_{t}$ estimator the bias was less than $1.6014 (\times 10^{-2})$ and the MSE was less than $2.4257(\times 10^{-3})$. The optimal cutoff estimator has bias less than $1.1685 (\times 10^{-2})$ and MSE $2.9361 (\times 10^{-3})$.

 The proposed estimators produce reasonably small biases and MSEs. As expected, both bias and MSE decrease with increasing sample size, thereby confirming the consistency of the estimators. Moreover, the empirical SD and ASD are very close. Both SDs increase with an increase in censoring rate and decrease with an increase in sample size.

Overall coverage of the confidence intervals for $\widehat{\mathrm{TPAUC}}_{t}$  estimator was close to the nominal $95\%$ .

\begin{table}[h]
\centering
\caption{Simulation results for the $\mathrm{PJ}_t$ and $m^{\mathrm{opt}}_t$ estimators: Bias ($\times$100), MSE ($\times$1000), SD, for different sample sizes ($n$), censoring rates (cens), and prediction times ($t$).}
\vspace{2mm}
\small
\label{T2}
\setlength{\tabcolsep}{10pt}
\renewcommand{\arraystretch}{2}
\begin{tabular}{ccc|ccc|ccc}
\hline
\multirow{2}{*}{$n$}
& \multirow{2}{*}{cens}
& \multirow{2}{*}{$t$}
& \multicolumn{3}{c|}{$\mathrm{PJ}_{t}$}
& \multicolumn{3}{c}{$m^{\mathrm{opt}}_t$}
\\[1.5mm]

\hhline{~~~|---|---}

&
&
& \rule{0pt}{4mm}SD
& Bias$(\times 10^2)$
& MSE$(\times 10^3)$
& SD
& Bias$(\times 10^2)$
& MSE$(\times 10^3)$
\\

\hline

\multicolumn{9}{l}{$\beta = -0.5$} \\

\multirow{6}{*}{500}
& 35 & 0.9 & 0.0299 & 0.9237 & 0.9795 & 0.0336 & 0.3837 & 1.1445 \\
& 35 & 1.2 &0.0314 & 0.8884 & 1.0620& 0.0356 & 0.0698 & 1.2676 \\
& 35 & 1.6 &0.0345 & 1.1940 & 1.3297 & 0.0382 & 0.0345 & 1.4552 \\
& 50 & 0.9 & 0.0319 & 0.5442 & 1.0456 & 0.0343 & 0.2080 & 1.1798 \\
& 50 & 1.2 & 0.0339 & 0.5762 & 1.1779 & 0.0361 & 0.08699 & 1.3024 \\
& 50 & 1.6 &0.0391 & 0.8819 & 1.6016& 0.0406 & 0.0935 & 1.6502 \\

\cline{2-9}

\multirow{6}{*}{1000}
& 35 & 0.9 &0.0215 & 0.656 & 0.5045 &0.0261 & 0.1698&0.6809 \\
& 35 & 1.2 &0.0226 &0.5621 &0.5399 & 0.0283& 0.2966& 0.8112\\
& 35 & 1.6 &0.0252 & 1.014& 0.7386&0.0311 & 0.1129&0.9648 \\
& 50 & 0.9 &0.0222 &0.4898 &0.5171 & 0.0266 & 0.1495& 0.7074\\
& 50 & 1.2 & 0.0241&0.3740 &0.5956 &0.0291 &0.1888 & 0.8514 \\
& 50 & 1.6 &0.0280 &0.7940 & 0.8454&0.0320 &0.0050 & 1.0257\\

\hline

\multicolumn{9}{l}{$\beta = -0.75$} \\

\multirow{6}{*}{500}
& 35 & 0.9 &0.0364 & 1.5053 & 1.5496 & 0.0449 & 0.8829 & 2.0899 \\
& 35 & 1.2 & 0.0369 & 1.5396 & 1.5962 & 0.0463 & 0.0534 & 2.1456 \\
& 35 & 1.6 &0.0418 & 1.6014 & 2.0039& 0.0512 & 1.0600 & 2.7343 \\
& 50 & 0.9 & 0.0389 & 1.2170 & 1.6630 & 0.0444 & 0.9722 & 2.0660 \\
& 50 & 1.2 & 0.0399 & 1.2257 & 1.7441 & 0.0473 & 0.0200 & 2.2334 \\
& 50 & 1.6 & 0.0473 & 1.3871 & 2.4257 & 0.0529 & 1.1685 & 2.9361 \\

\cline{2-9}

\multirow{6}{*}{1000}
& 35 & 0.9 &0.0253  & 0.9836&0.7360 &0.0355 & 1.0117& 1.3633 \\
& 35 & 1.2 &0.0270 & 1.1675& 0.8640&0.0376 & 0.1701 & 1.4157\\
& 35 & 1.6 & 0.0292& 1.3574 & 1.034 & 0.0407 &1.0957 &1.775 \\
& 50 & 0.9 &0.0277 &0.8091 &0.8320 & 0.0353& 0.9928& 1.343 \\
& 50 & 1.2 & 0.0294 & 1.007& 0.9669&0.0389 & 0.0903 & 1.510\\
& 50 & 1.6 & 0.0329& 1.137 & 1.213& 0.0429&0.9625 &1.929 \\

\bottomrule
\end{tabular}
\end{table}

%%%%%%%%%%%%%%%%%%%%%%    simulation tables end    %%%%%%%%%%%%%%%%%%%%%%

.

%%%%%%%%%%%%%%%%%%%%%%%%%%%%   Real data  %%%%%%%%%%%%%%%%%%%%%%%%%%%%%%%%

\section{Real Data Applications}

In this section, we illustrate our proposed method using two real-world datasets. The first dataset is from the Primary Biliary Cirrhosis study, and the latter is obtained from the Molecular Taxonomy of Breast Cancer International Consortium trial.

\subsection{An application to Primary Biliary Cirrhosis data}

The Primary Biliary Cirrhosis (PBC) dataset was obtained from a clinical study conducted at the Mayo Clinic between 1974 and 1984. The data consists of 312 patients diagnosed with PBC, a rare autoimmune liver disease. Patients were randomly assigned to one of two treatment groups: a placebo group and a D-penicillamine treatment group. During the follow-up period, 125 patients died, while the remaining participants were right-censored. The survival time ranges from $0.11$ to $12.50$ years.

Heagerty et al.\cite{Heagertyzheng2005} developed two prognostic biomarkers using the Cox proportional hazards model to illustrate their proposed time-dependent ROC methodology. The first biomarker (Biomarker~1) was constructed using five covariates: $\log(\text{bilirubin})$, albumin, $\log(\text{prothrombin time})$, edema, and age. The second biomarker (Biomarker~2) was derived from the same set of covariates, excluding $\log(\text{bilirubin})$. A detailed summary of these covariates are available in reference\cite{Heagertyzheng2005}. The dataset, along with the two derived biomarkers, is publicly available in the R-package \texttt{survivalROC}.\cite{heagertyrpack} Our aim is to evaluate the predictive performance of these two biomarkers for different restrictions of TPR and FPR at different time horizons.

As discussed in the introduction, for a reliable medical diagnostic test, it is crucial to achieve a TPR that exceeds 50\%; therefore, the lower bound of the TPR is set at 0.5. On the other hand, a diagnostic test yielding more than 50\% false positives is generally unacceptable in medical applications. In fact, for certain critical diseases, the maximum tolerable FPR is often restricted to 15\%. Consequently, we consider two upper bounds, 0.5 and 0.15, for the FPR. Two time horizons of 2 years and 4 years are assumed in the following discussion. The estimated values of $\mathrm{TPAUC}_{t}$, $\mathrm{PJ}_{t}$ , optimal cut off based on the proposed estimators and 95\% confidence interval based on 1000 bootstrap samples are reported in Table \ref{Tpbc}.

\begin{figure}[!ht]
    \centering
    % Define the width for each subfigure. 
    % A value of 0.49\textwidth ensures four images fit nicely in a 2x2 grid 
    % with a little space between them.
    \begin{subfigure}[b]{0.4\textwidth}
     \centering
        \includegraphics[width=\textwidth]{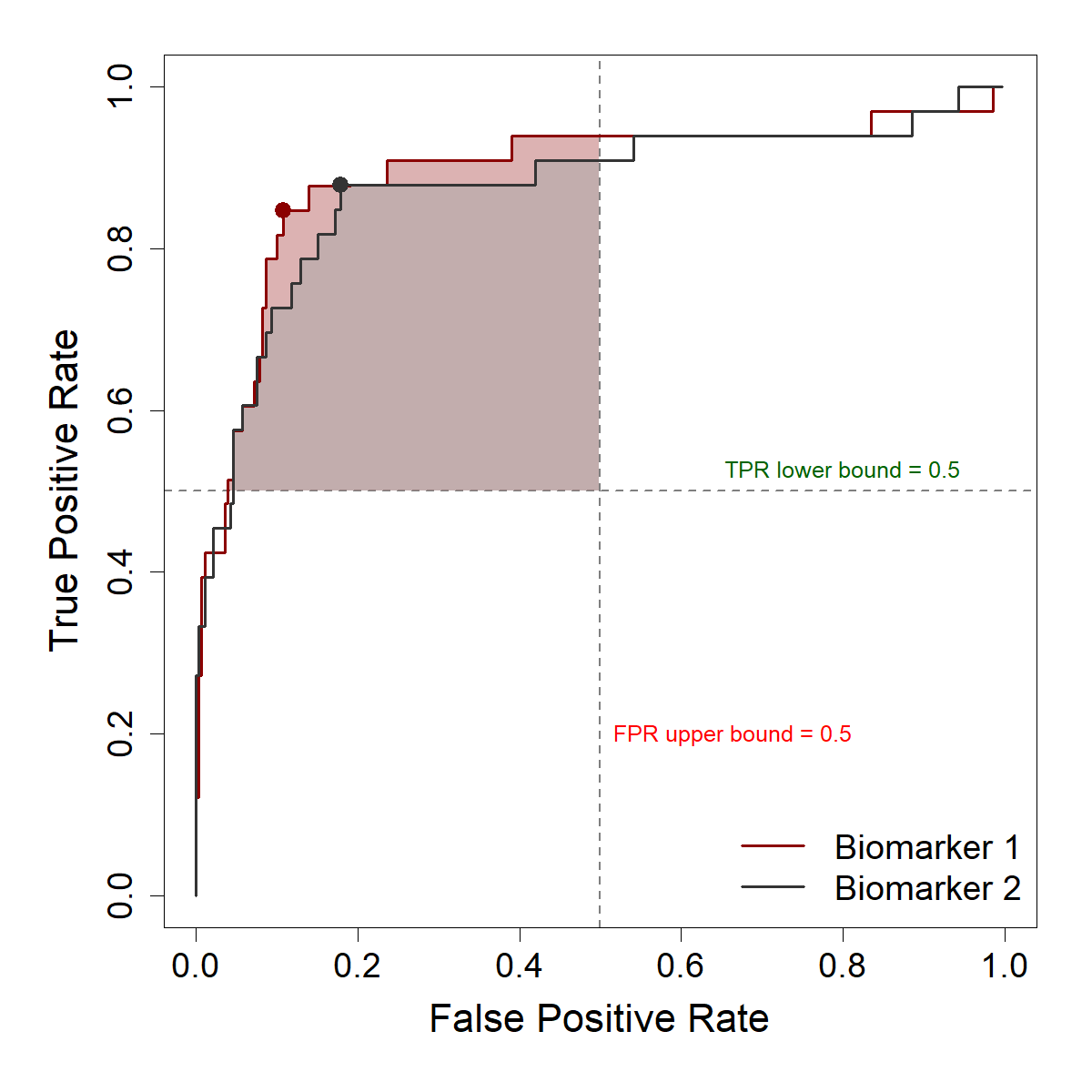}
        \label{fig:plotA}
    \end{subfigure}
    \begin{subfigure}[b]{0.4\textwidth}
        \centering
        \includegraphics[width=\textwidth]{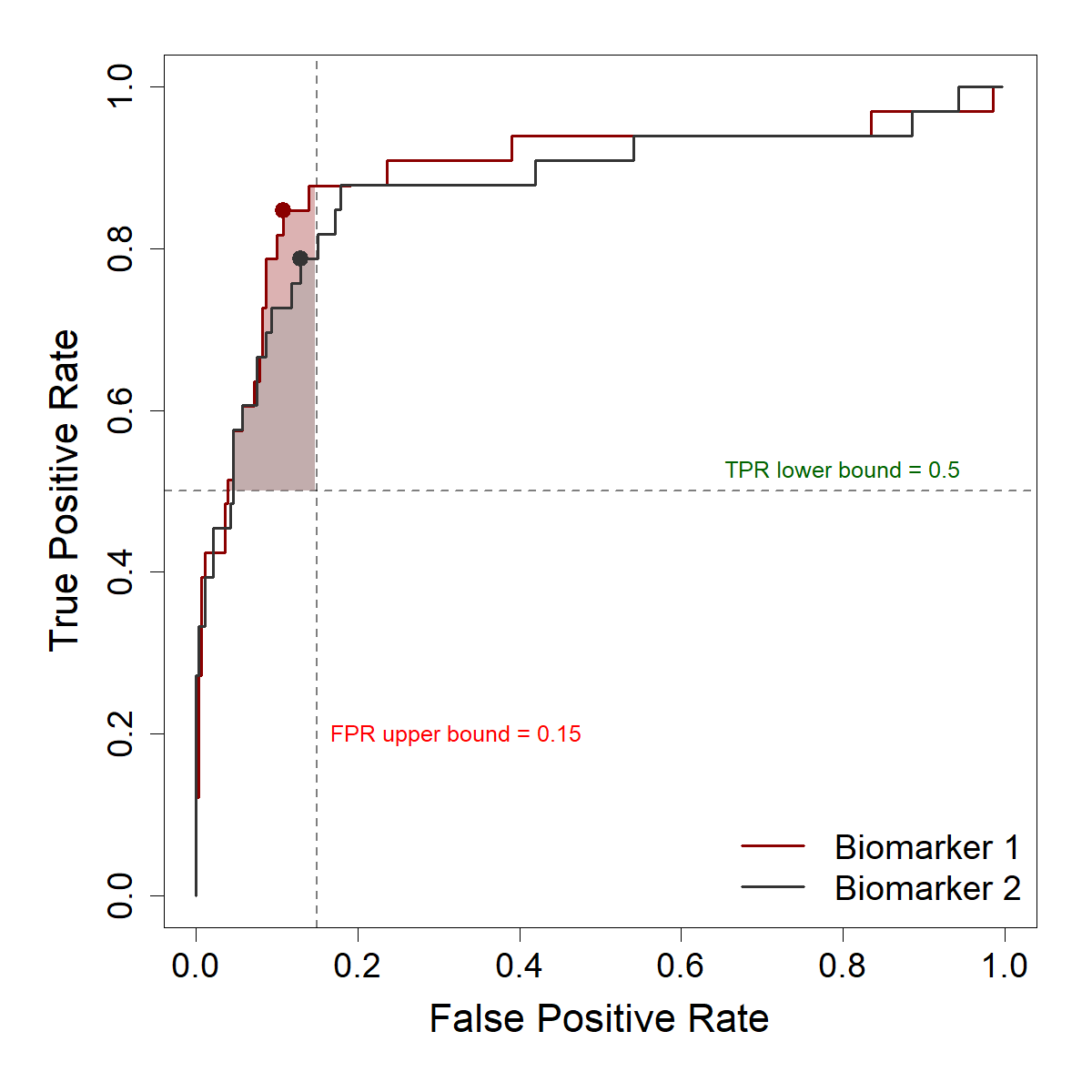}
        \label{fig:plotB}
    \end{subfigure}

    % Force a line break after the first row
    \par\bigskip

    \begin{subfigure}[b]{0.4\textwidth}
        \centering
        \includegraphics[width=\textwidth]{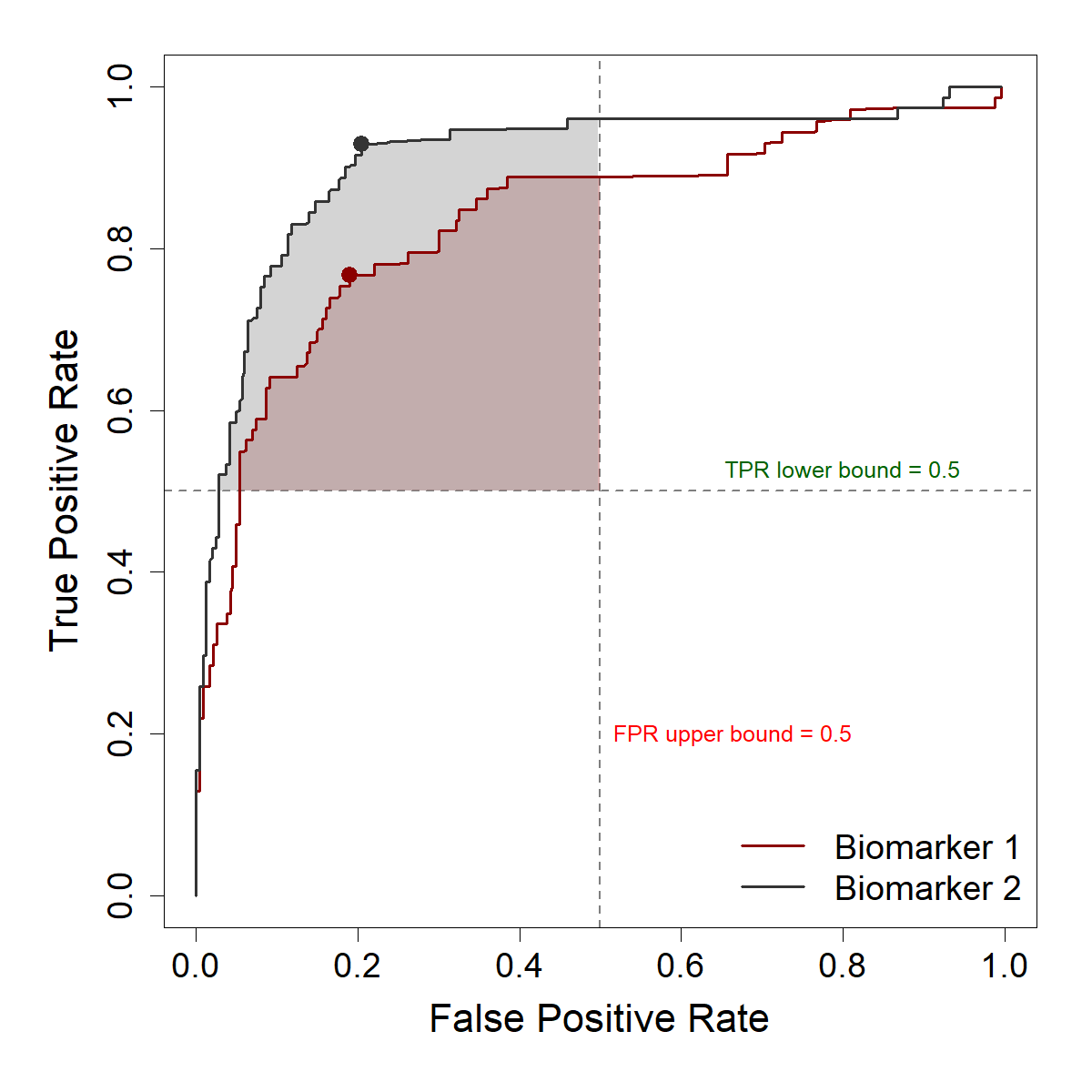}
        \label{fig:plotC}
    \end{subfigure}
    \begin{subfigure}[b]{0.4\textwidth}
        \centering
        \includegraphics[width=\textwidth]{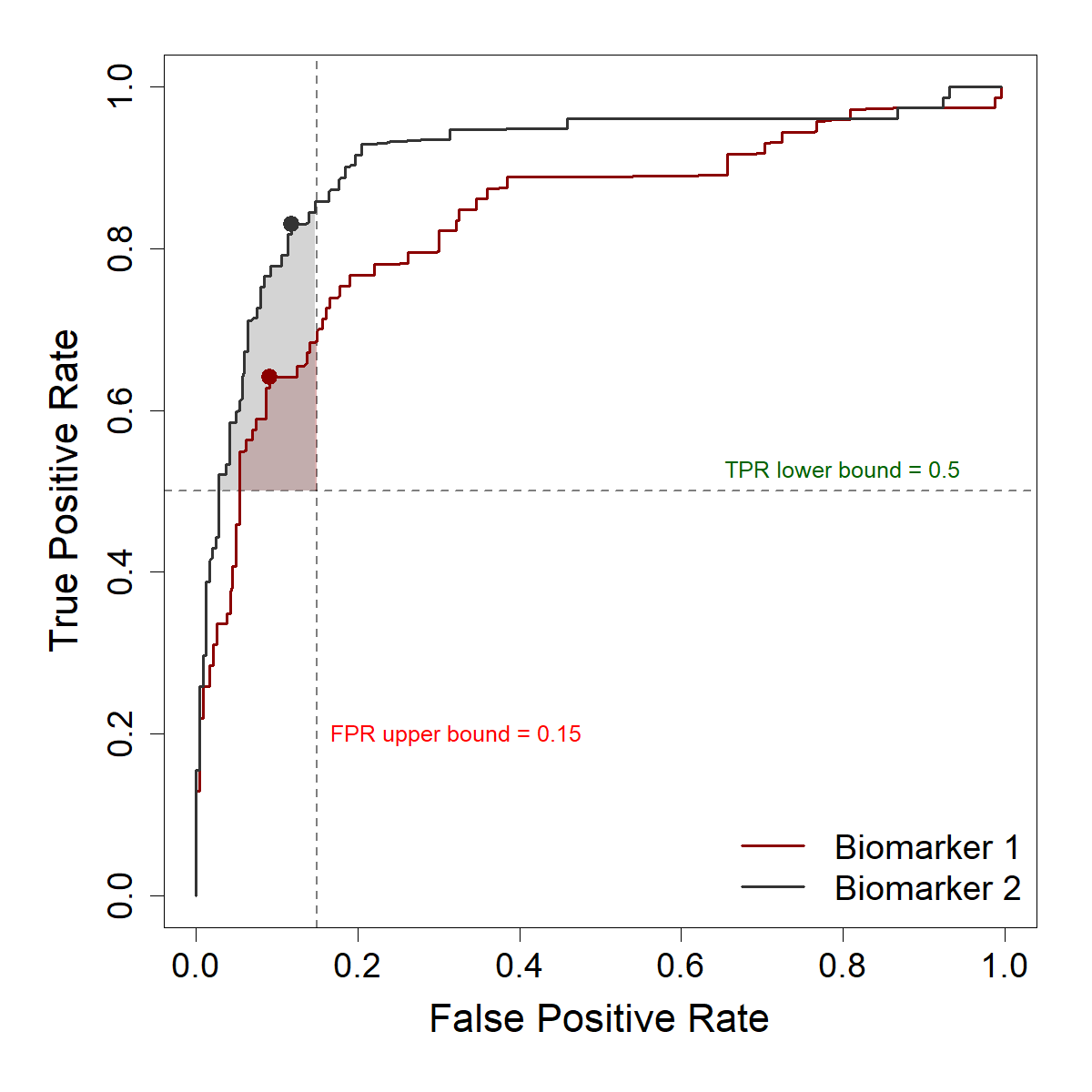}
        \label{fig:plotD}
    \end{subfigure}

    % Overall caption for the entire 2x2 grid
    \caption{The estimated TPAUC and partial Youden index point based on PBC dataset for different restrictions of TPR lower bound and FPR upper bound. The first column correspond to $l_o=0.5$, $u_0=0.5$ and the second column correspond to $l_o=0.5$, $u_0=0.15$. The first and second rows correspond to time horizons $t=2$ and $t=4$ years, respectively.}
    \label{fig:four_plots_pbc}
\end{figure}

Figure~\ref{fig:four_plots_pbc} shows the estimated ROC curves for Biomarker~1 and Biomarker~2, together with the corresponding estimated two-way partial area under the ROC curve and the estimated partial Youden index points, for different time horizons and different TPR and FPR bounds. The figures in the first row correspond to the time horizon of 2 years, and the second row corresponds to the time horizon of 4 years. The estimated ROC curves for Biomarker~1 and Biomarker~2 are shown in red and green, respectively. For each time horizon, the figures in the left panel correspond to a TPR lower bound of 50\% and an FPR upper bound of 50\%, while the figures in the right panel correspond to the same TPR lower bound of 50\% and an FPR upper bound of 15\%.

From Table \ref{Tpbc} and Figure~\ref{fig:four_plots_pbc}, we observe that the estimates of TPAUC for different restrictions on TPR and FPR reveal that Biomarker~1 shows a better overall prognostic ability over Biomarker~2 for a 2-year time horizon. However, when the time horizon is 4 years, Marker~2 demonstrates improved prognostic ability over the region of interest relative to Marker~1. This reversal in the performance of biomarkers in terms of prognostic ability over time underscores the critical importance of incorporating a time dimension into ROC analysis. We also note that the $\mathrm{PJ}_{t}$ and the corresponding optimal cutoff values are also changed when the FPR upper bound is reduced to 15\% from 50\%  highlighting a more accurate assessment of the highest prognostic accuracy achievable by the biomarker in the region of clinical interest. Similar to TPAUC, these values also change when the time horizon is changed from $2$ years to $4$ years.

\begin{table}[!ht]
\centering
\setlength{\tabcolsep}{5pt} % column spacing
\renewcommand{\arraystretch}{1.2} % row spacing
\caption{Estimated time-dependent TPAUCs, partial Youden indices and the corresponding cutoff points for two biomarkers at different time horizons with different TPR and FPR bounds in PBC dataset.}
\begin{tabular}{l l | c c c c c c}
\toprule
\label{Tpbc}
\multirow{2}{*}{$t$} &
\multirow{2}{*}{Biomarker} & 
\multicolumn{2}{c}{$\text{TPAUC}_{t}$} & 
\multicolumn{2}{c}{$\mathrm{PJ}_{t}$} & 
\multicolumn{2}{c}{$m^{\mathrm{opt}}_t$} \\
\cmidrule(lr){3-4} \cmidrule(lr){5-6} \cmidrule(lr){7-8}
& & Estimate & CI & Estimate & CI & Estimate & CI \\
\midrule
\multicolumn{8}{l}{\small \textit{$l_0 = 0.5$, $u_0 = 0.5$}} \\
\addlinespace[2pt]
\multirow{2}{*}{$2$} & Biomarker 1 & 0.164 & [0.109, 0.202] & 0.740 & [0.633, 0.870] & 7.268 & [6.997, 7.504] \\
                       & Biomarker 2 & 0.148  & [0.086, 0.189] &   0.699 & [0.582, 0.821] & 6.911 &[6.898, 7.514]\\
\cmidrule(lr){1-8}
\multirow{2}{*}{$4$} & Biomarker 1 & 0.124 & [0.084, 0.160] & 0.576 & [0.478, 0.697] & 6.682 & [6.321, 7.059] \\
                       & Biomarker 2 & 0.177 & [0.144, 0.206]  & 0.724 & [0.661, 0.819] &  6.226& [6.177, 6.911] \\
\midrule
\multicolumn{8}{l}{\small \textit{$l_0 = 0.5$, $u_0 = 0.15$}} \\
\addlinespace[2pt]
\multirow{2}{*}{$2$} & Biomarker 1 & 0.025 & [0.008, 0.042] & 0.740 & [0.611, 0.866] & 7.268 & [7.057, 7.504] \\
                       & Biomarker 2 &0.019  & [0.004, 0.036] & 0.658 & [0.517, 0.808] & 7.249 & [6.911, 7.644] \\
\cmidrule(lr){1-8}
\multirow{2}{*}{$4$} & Biomarker 1 & 0.011 & [0.002, 0.024] & 0.550 & [0.448, 0.686] &7.035  & [6.700, 7.268] \\
                       & Biomarker 2 & 0.027 & [0.014, 0.043] & 0.712 & [0.621, 0.816] & 6.628 & [6.295, 6.977] \\
\bottomrule
\end{tabular}
\end{table}

\subsection{An application to METABRIC data}

In this section, we illustrate our proposed method using a breast cancer dataset from the Molecular Taxonomy of Breast Cancer International Consortium (METABRIC) study. These cohorts were collected from five tumor banks in the United Kingdom and Canada.\cite{curtis2012genomic,mukherjee2018associations} We obtained the dataset from \href{https://www.cbioportal.org/study/summary?id=brca_metabric}{cBioPortal}.

Typically, a breast cancer diagnosed patient undergoes a combination of treatment interventions, including surgical interventions, hormone therapy, radiotherapy, and chemotherapy. After removing the records with missing values, the dataset contains information on the survival time (in months) following these treatment interventions for 1092 patients. Observed survival times range from 0.10 to 337.03 months, with a median duration of 116 months. At the end of the study, 608 patients died, and the remaining patients were right censored.

As in \cite{Heagertyzheng2005,Beyene2022Time-dependentData,Dey2023InferenceMeasures}  we derive two prognostic risk scores (or biomarkers) using the Cox proportional hazards model. We consider the following clinical covariates: age at diagnosis, tumour size, tumour stage, number of positive lymph nodes examined, Nottingham Prognostic Index (NPI), Hormone therapy, radiotherapy, chemotherapy, and Human epidermal growth factor receptor status (HER2 status). Table~\ref{tab: summarycovex2} contains a summary of these covariates. Among these covariates, the NPI is a significant prognostic factor in breast cancer studies. It is estimated using tumour size, tumour stage and number of positive lymph nodes.We derive two biomarkers. the first biomarker (Biomarker 1) was constructed using all covariates except NPI. To calculate this biomarker, we dictomize the tumor stage by considering all patients with tumor stage I and II into one category and rest in the other. The second biomarker (Biomarker 2) was obtained using the same set of covariates, with NPI replacing tumour size, tumour stage, and the number of positive lymph nodes. The biomarkers are the filled linear predictors from the corresponding cox models. The two biomarkers, along with their constituent covariates and corresponding Cox model estimates, are summarised in Table~\ref{tab:cox_combined}. Our aim here is to assess the prognostic ability of these biomarkers using our proposed estimators. 

\begin{table}[ht]
\centering
\caption{Summary statistics of clinical covariates present in the METABRIC study}
\label{tab:clinical_summary}
\begin{tabular}{ll}
\toprule
\textbf{Covariate} & \textbf{Summary} \\
\midrule
Age at diagnosis & 21.9 -- 96.3 years (median: 61 years) \\

Tumour size & 1 -- 180 mm (median: 22 mm) \\

Tumour stage & 
Stage I: 369 patients \\
& Stage II: 624 patients \\
& Stage III: 92 patients \\
& Stage IV: 7 patients \\

Number of positive lymph nodes & 0 -- 41 \\

Nottingham Prognostic Index (NPI) & 2 -- 6.36 (median: 4.05) \\

Hormone therapy & 670 patients \\

Radiotherapy & 725 patients \\

Chemotherapy & 241 patients \\

HER2 status & 958 negative, 148 positive \\
\bottomrule
\end{tabular}
\label{tab: summarycovex2}
\end{table}

Similar to the previous example, we considered two pairs of TPR lower bound and FPR upper bound, namely $(l_0=0.5, u_0=0.5)$ and $(l_0=0.5, u_0=0.3)$. Two time horizons of 116 months $(t_1)$ and 180 months $(t_2)$ were chosen arbitarily. The estimated values of ${\mathrm{TPAUC}}_{t}$, ${\mathrm{PJ}}_{t}$, optimal cutoff $m^{\mathrm{opt}}_t$ based on the proposed estimators and 95\% confidence interval based on 1000 bootstrap samples are reported in Table \ref{tabmetabric}.

\begin{figure}[H]
    \centering
    % Define the width for each subfigure. 
    % A value of 0.49\textwidth ensures four images fit nicely in a 2x2 grid 
    % with a little space between them.
    \begin{subfigure}[b]{0.4\textwidth}
     \centering
        \includegraphics[width=\textwidth]{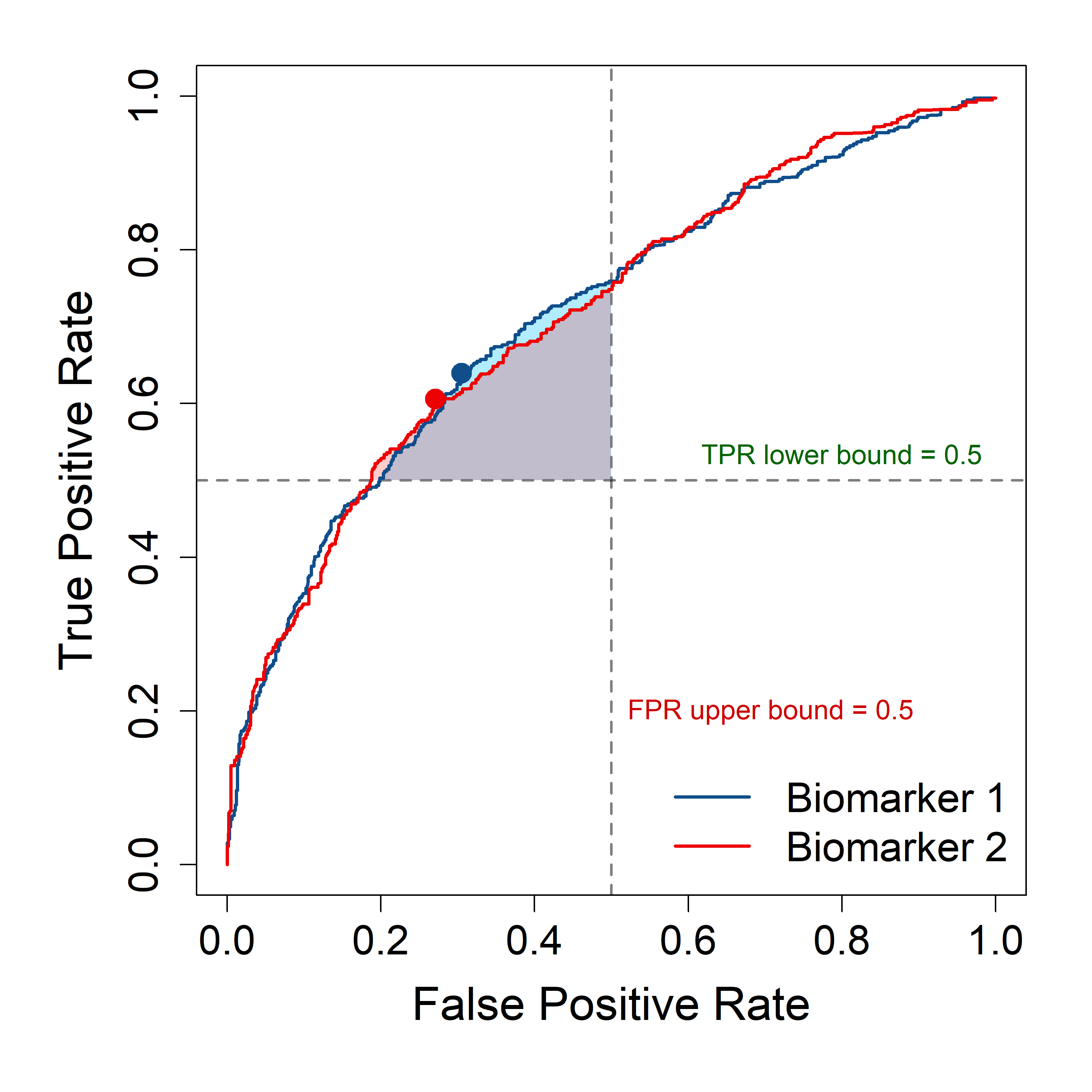}
        \label{fig:plotA1}
    \end{subfigure}
    \begin{subfigure}[b]{0.4\textwidth}
        \centering
        \includegraphics[width=\textwidth]{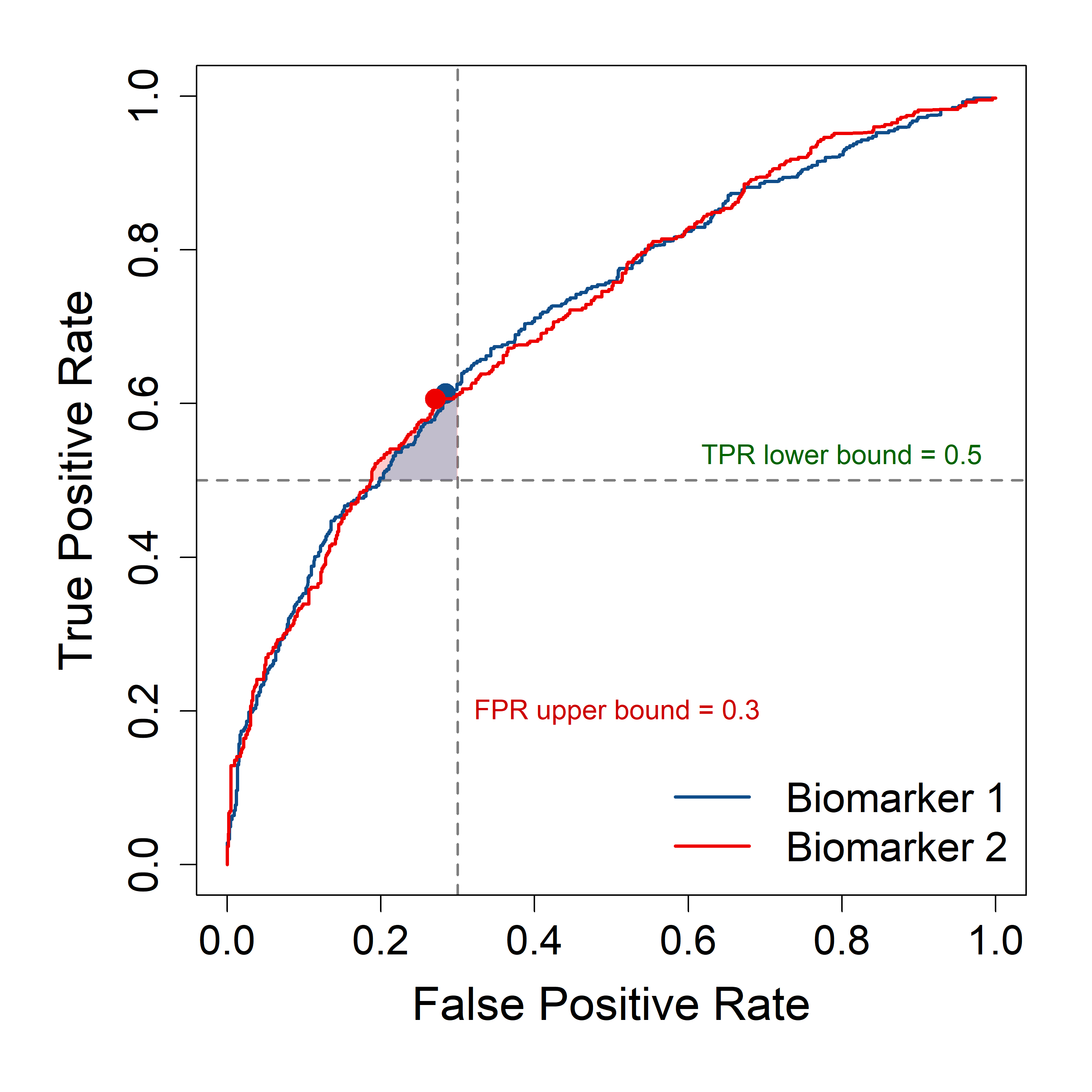}
        \label{fig:plotB1}
    \end{subfigure}

    % Force a line break after the first row
    \par\bigskip

    \begin{subfigure}[b]{0.4\textwidth}
        \centering
        \includegraphics[width=\textwidth]{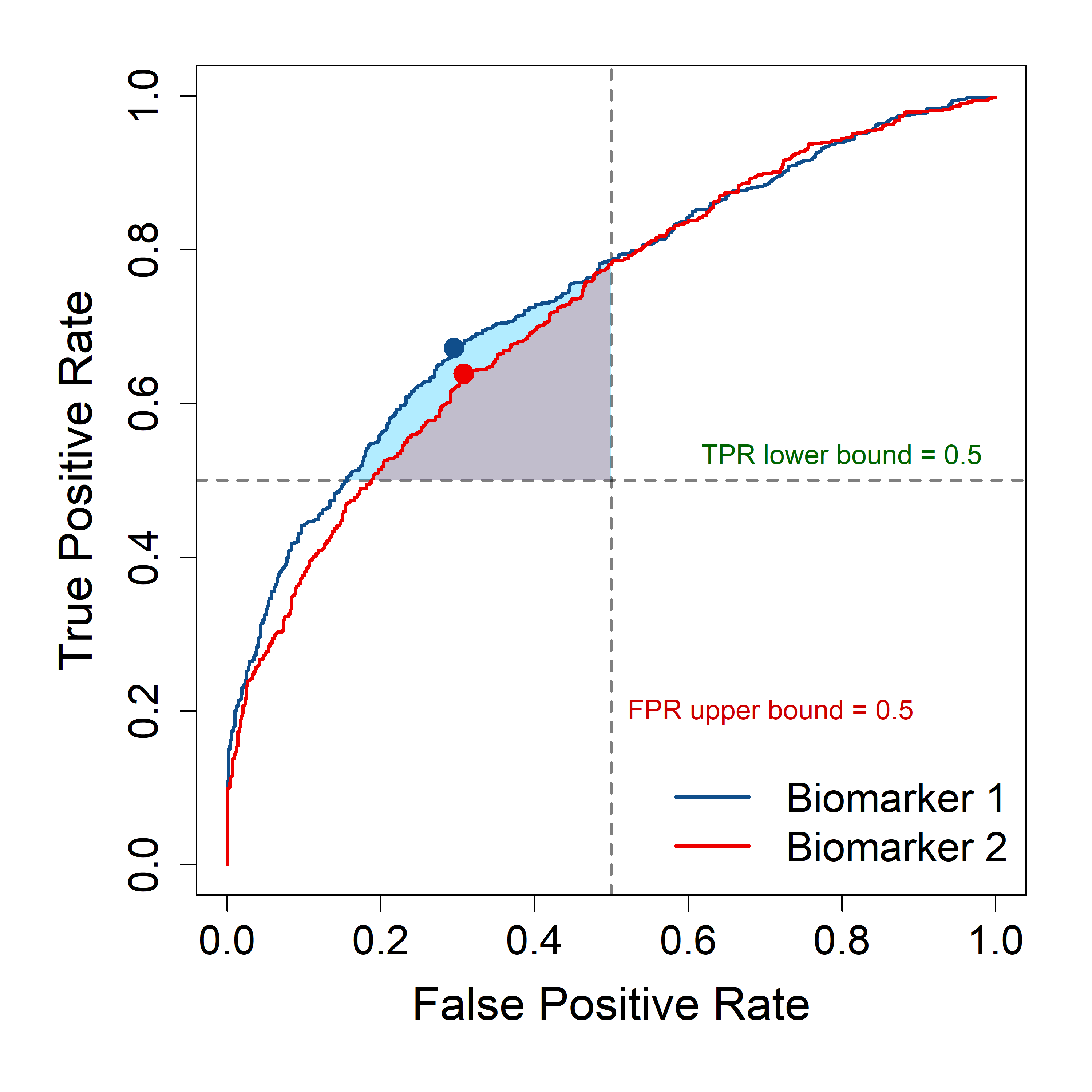}
        \label{fig:plotC1}
    \end{subfigure}
    \begin{subfigure}[b]{0.4\textwidth}
        \centering
        \includegraphics[width=\textwidth]{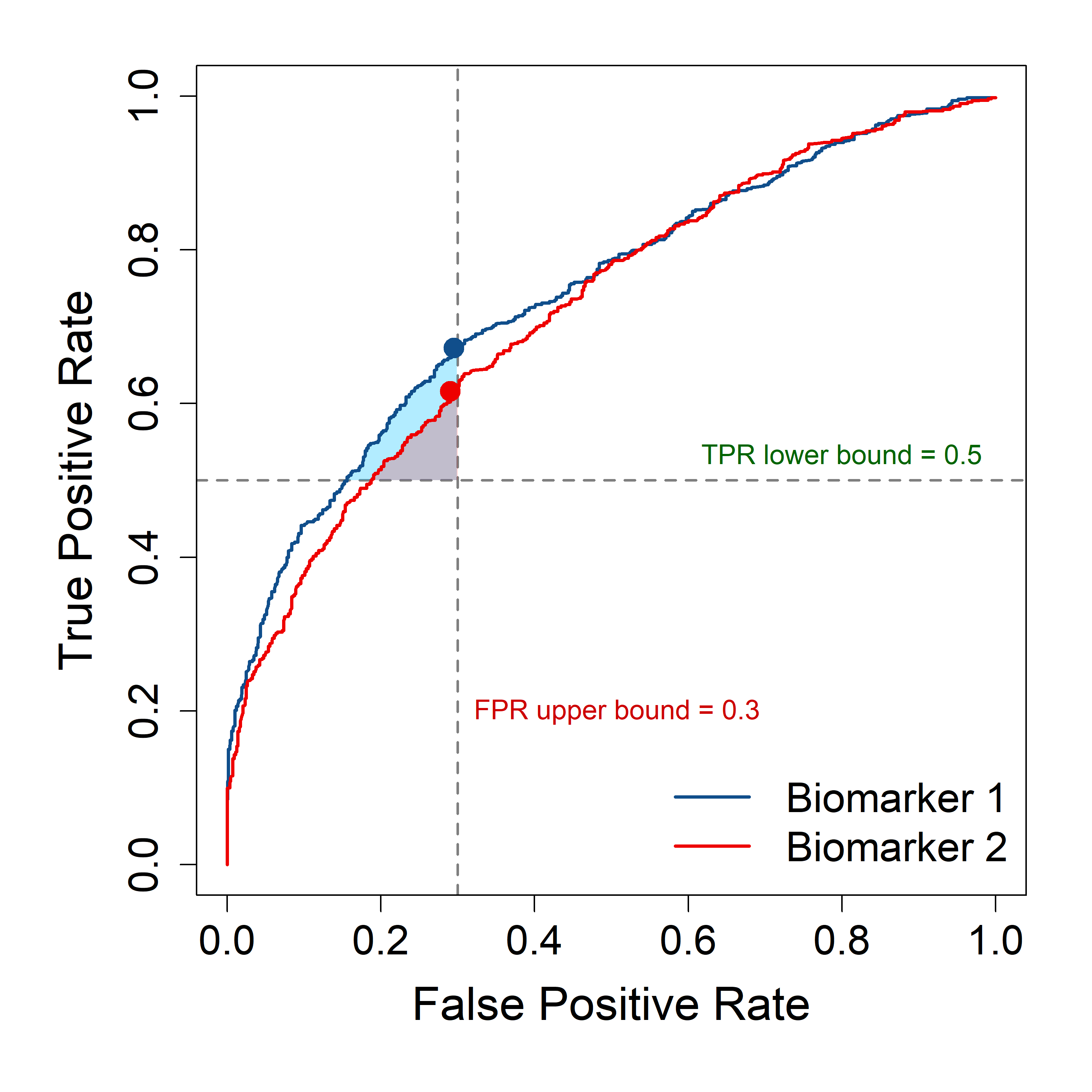}
        \label{fig:plotD1}
    \end{subfigure}

    % Overall caption for the entire 2x2 grid
    \caption{The estimated TPAUC and partial Youden index point based on METABRIC dataset for different restrictions of TPR lower bound and FPR upper bound, respectively. The first column correspond to $l_o=0.5$, $u_0=0.5$ and the second column correspond to $l_o=0.5$, $u_0=0.3$. The first row corresponds to time horizon 116 months, and the second row corresponds to time horizon 180 months.}
    \label{fig:four_plots_METABRIC_S}
\end{figure}

The Figure~\ref{fig:four_plots_METABRIC_S} shows the estimated ROC curves for Biomarker~1 and Biomarker~2, together with the corresponding estimated TPAUC and the estimated partial Youden index points, for different time horizons and different TPR and FPR bounds. The figures in the first row correspond to the time horizon $t_1=116$ months and the second row corresponds to the time horizon $t_2=180$ months. The estimated ROC curves for Biomarker~1 and Biomarker~2 are shown in blue and red, respectively. For each time horizon, the figures in the left panel correspond to a TPR lower bound of 50\% and a FPR upper bound of 50\%, while the figures in the right panel correspond to the same TPR lower bound of 50\% and a FPR upper bound of 30\%.

From Table~\ref{tabmetabric} and  Figure~\ref{fig:four_plots_METABRIC_S}, we observe that for the shorter time horizon 116 months, the estimate of ${\mathrm{TPAUC}}_{t}$ indicates that Biomarker~1 performs better than Biomarker~2 when the region of interest is defined by a TPR lower bound of 50\% and an FPR upper bound of 50\%. However, when the region is made more stringent by reducing the FPR upper bound from 50\% to 30\%, Biomarker~2 demonstrates better prognostic performance than Biomarker 1. For $\mathrm{PJ}_{t}$, we note that for FPR upper bound of 50\%, the $\mathrm{PJ}_{t}$ value of Biomarker~1 is very close to that of  Biomarker~2; however, when the FPR upper bound is reduced to 30\%, then the $\mathrm{PJ}_{t}$ value of Biomarker~2 is more than Biomarker~1. Indicating the highest achievable prognostic ability of Biomarker~2 will be more than the highest achievable prognostic ability of Biomarker~1 for a stricter FPR upper bound.  

When the time horizon is extended to $180$ months, Biomarker~1 consistently outperforms Biomarker~2 across both sets of TPR/FPR bounds. This suggests that, for shorter follow-up time, if a lower false-positive rate is desired, then the biomarker incorporating the NPI shows better overall prognostic ability than the biomarker based on individual tumour characteristics (tumour size, tumour stage, and number of positive lymph nodes), given the same set of other clinical covariates. A similar situation has also been indicated by $\mathrm{PJ}_{t}$ for the highest achievable prognostic ability of two biomarkers. This highlights the importance of incorporating time dependence and two-way bounds on TPR and FPR when evaluating the prognostic performance of biomarkers.

\begin{table}[htbp]
\centering
\caption{Cox proportional hazards model estimates for Biomarker 1 (without NPI) and Biomarker 2 (with NPI) based on the METABRIC dataset.}
\label{tab:cox_combined}
\setlength{\tabcolsep}{14pt}
\renewcommand{\arraystretch}{1.2}
\renewcommand{\arraystretch}{1.5} % row spacing
% ============================
% Model 1
% ============================
\begin{tabular}{lcccc}
\hline
\multicolumn{5}{c}{\textbf{Biomarker 1 (Without NPI)}} \\
\hline
\textbf{Covariates} & $\widehat{\beta}$ & SE & $z-score$ & $p$-value \\
\hline
Age at Diagnosis                   & 0.04286 & 0.00408 & 10.5147 & $7.39\times10^{-26}$ \\
Tumor Size                         & 0.00918 & 0.00238 & 3.8529  & $1.17\times10^{-4}$ \\
Tumor Stage                        & 0.02495 & 0.16124 & 0.1547  & $8.78 \times10^{-1}$ \\
Lymph Nodes Positive               & 0.05825 & 0.01083 & 5.3782  & $7.52\times10^{-8}$ \\
Breast Surgery: Mastectomy         & 0.11227 & 0.10925 & 1.0276  & $3.04\times10^{-1}$\\
Hormone Therapy (Yes)              & $-0.13408$ & 0.09004 & $-1.4890$ &$1.36\times10^{-1}$\\
Radiotherapy (Yes)                 & $-0.21270$ & 0.10809 & $-1.9677$ &$4.91\times10^{-2}$ \\
Chemotherapy (Yes)                 & 0.40126 & 0.13199 & 3.0401  & $2.36\times10^{-3}$ \\
HER2 Positive                      & 0.55788 & 0.12169 & 4.5844  & $4.55\times10^{-6}$ \\
\hline
\end{tabular}

\vspace{0.3cm}

% ============================
% Model 2
% ============================
\begin{tabular}{lcccc}
\hline
\multicolumn{5}{c}{\textbf{Biomarker 2 (With NPI)}} \\
\hline
\textbf{Covariates} & $\widehat{\beta}$ & SE & $z-score$ & $p$-value \\
\hline
Age at Diagnosis                   & 0.04492 & 0.00410 & 10.9604 & $5.92\times10^{-28}$ \\
Breast Surgery: Mastectomy         & 0.17702 & 0.10549 & 1.6781  & $9.33\times10^{-2}$ \\
Nottingham Prognostic Index        & 0.32452 & 0.05092 & 6.3737  & $1.85\times10^{-10}$ \\
Hormone Therapy (Yes)              & $-0.19169$ & 0.09328 & $-2.0551$ & $3.98\times10^{-2}$ \\
Radiotherapy (Yes)                 & $-0.22133$ & 0.10712 & $-2.0615$ & $3.88\times10^{-2}$ \\
Chemotherapy (Yes)                 & 0.36603 & 0.13482 & 2.7149  & $6.63\times10^{-3}$ \\
HER2 Positive                      & 0.45891 & 0.12058 & 3.8057  & $1.44\times10^{-4}$ \\
\hline
\end{tabular}
\end{table}

\begin{table}[ht]
\centering
\setlength{\tabcolsep}{5pt} % column spacing
\renewcommand{\arraystretch}{1.4} % row spacing
\caption{Estimated time-dependent TPAUCs, partial Youden indices and the corresponding cutoff points for two biomarkers at different time horizons with different TPR and FPR bounds based on METABRIC dataset.}
\begin{tabular}{l l | c c c c c c}
\toprule
\multirow{2}{*}{$t$} &
\multirow{2}{*}{Biomarker} & 
\multicolumn{2}{c}{$\text{TPAUC}_{t}$} & 
\multicolumn{2}{c}{$\mathrm{PJ}_{t}$} & 
\multicolumn{2}{c}{$m^{\mathrm{opt}}_t$} \\
\cmidrule(lr){3-4} \cmidrule(lr){5-6} \cmidrule(lr){7-8}
& & Estimate & CI & Estimate & CI & Estimate & CI \\
\midrule
\multicolumn{8}{l}{\small \textit{$l_0 = 0.5$, $u_0 = 0.5$}} \\
\addlinespace[2pt]
\multirow{2}{*}{$t_1$} & Biomarker 1 & 0.047 & [0.033, 0.063] &  0.334& [0.281, 0.399] & 0.092 & $[-0.05, 0.327]$ \\
                       & Biomarker 2 & 0.044 & [0.030, 0.060] & 0.335 & [0.286, 0.398] & 0.137 & [0.126, 0.295] \\
\cmidrule(lr){1-8}
\multirow{2}{*}{$t_2$} & Biomarker 1 & 0.059 & [0.042, 0.076] & 0.377 & [0.324, 0.446] & $-0.022$ & $[-0.064, 0.168]$ \\
                       & Biomarker 2 & 0.046 & [0.031, 0.063] &  0.331 & [0.277, 0.393]
 & $-0.005$ & $[-0.150, 0.249]$
 \\
\midrule
\multicolumn{8}{l}{\small \textit{$l_0 = 0.5$, $u_0 = 0.3$}} \\
\addlinespace[2pt]
\multirow{2}{*}{$t_1$} & Biomarker 1 & 0.006 & [0.002, 0.014] & 0.328 & [0.271, 0.393] & 0.129 &[0.065, 0.341]  \\
                       & Biomarker 2 & 0.008 & [0.002, 0.015] & 0.335 & [0.288, 0.399] & 0.137 & [0.133, 0.295] \\
\cmidrule(lr){1-8}
\multirow{2}{*}{$t_2$} & Biomarker 1 & 0.013 & [0.006, 0.023] & 0.377 & [0.325, 0.444] & $-0.022$ &$[-0.045, 0.175]$\\
                       & Biomarker 2 & 0.007 & [0.002, 0.014]& 0.325 & [0.273, 0.39] & 0.029 &$[-0.006, 0.249]$\\
\bottomrule
\end{tabular}
\label{tabmetabric}
\end{table}

\section{Discussion and conclusion}
ROC analysis is a widely used tool for assessing the diagnostic and prognostic ability of a candidate biomarker. Given the practical advantages of focusing on clinically meaningful regions of the ROC space, several studies have explored partial ROC analysis within the region of interest. Most of these literatures are based on studies in which the event status is known. However, in prognostic studies, such as time-to-event analyses, the event status is time-dependent, and some subjects are censored. The statistical literature on time-dependent ROC analysis for a region of clinical interest is limited. To address this research gap, this manuscript proposes TPAUC and partial Youden index estimators for cumulative/dynamic ROC curves in the presence of right-censored survival data. The proposed estimators are non-parametric in nature and thus make no parametric model assumptions. It also efficiently accounts for right censoring.

The simulation results demonstrated that the proposed estimators performed well, exhibiting negligible bias and MSE. Moreover, the MSE decreased with increasing sample size, indicating the consistency of the estimators. The ASD closely matched the SD, and the coverage probabilities of the constructed confidence intervals were generally close to the nominal $95\%$ level. Overall, these findings demonstrate the good performance of the proposed estimators and inference procedures.

 One limitation of the proposed method, as with other non-parametric methods, is its reliance on an adequate sample size. Moreover, imposing very strict bounds on TPR and FPR results in a substantial reduction of the effective sample size. This may potentially lead to an increase in bias. These limitations were also underscored by earlier authors in the context of time-dependent FPR-pAUC.\cite{hung2011nonparametric,jiang2024analyzing} Hence, for very strict TPR and FPR bounds, a moderate sample size is preferable to ensure an adequate effective sample size and reliable estimation.

 In this article, we have not discussed any statistical approach for selecting the optimal TPR lower bound and FPR upper bound. Hence, the method relies on practitioners' field experience to select suitable bounds for a candidate biomarker. However, in classical ROC analysis, few statistical approaches have been proposed for selecting such optimal bounds.\cite{lavazza2022considerations}. Many of these methods can be extended to time-dependent ROC. A comparative study of these approaches for time-dependent ROC analysis can be an interesting topic for future research. Finally, the proposed method is for a cumulative/dynamic time-dependent ROC. It would be an interesting future work to extend this to other time-dependent ROCs.

\bibliographystyle{unsrtnat}
 % or use abbrvnat, unsrtnat, etc.
\bibliography{Cite}   % references.bib file

% \appendix

\section{Appendix}
\subsection{Derivation of \texorpdfstring{$\mathrm{TPR}_t(m)$, $\mathrm{FPR}_t(m)$, $\mathrm{TPAUC}_t$, $\mathrm{PJ_t}$}{TPR, FPR, TPAUC, PJ}.} 
\label{app:dervation}

\begin{align}
\mathrm{TPR}_t(m) 
&= P(M > m \mid T \leq t) \notag \\[6pt]
&= \frac{E\{ I(M > m,\, T < t) \}}{E\{ I(T < t) \}} \notag \\[6pt]
&= \frac{E\{ I(M > m)\, E(I(T < t) \mid Y, \Delta, M) \}}
        {E\{ E(I(T < t) \mid Y, \Delta, M) \}} \notag\\[6pt]
    &= \frac{\sum_i I(M_i > m)\, {W}_{i1}(t) }
        {\sum_i {W}_{i1}(t) }  \notag  \quad [\text{Replacing the expectations by empirical average}]
    \end{align}
Replacing the weights by their corresponding estimates we get,
\begin{align}
        \widehat{\mathrm{TPR}}_{t}(m)= \frac{1 }{\widehat{N}_1(t)}\sum_{i=1}^n I(M_i > m)\widehat{W}_{i1}(t). 
\end{align}

\begin{align}
\mathrm{FPR}_t(m) 
&= P(M > m \mid T > t) \notag \\[6pt]
&= \frac{E\{ I(M > m,\, T > t) \}}{E\{ I(T > t) \}} \notag \\[6pt]
&= \frac{E\{ I(M > m)\, E(I(T > t) \mid Y, \Delta, M) \}}
        {E\{ E(I(T > t) \mid Y, \Delta, M) \}} \notag \\[6pt]
&= \frac{\sum_i I(M_i > m)\, {W}_{i0}(t) }
        {\sum_i {W}_{i0}(t) } \notag  \quad [\text{Replacing the expectations by empirical average}]
\end{align}
Replacing the weights by their corresponding estimates we get,
\begin{align}
            \widehat{\mathrm{FPR}}_{t}(m)= \frac{1}{\widehat{N}_0(t)}\sum_{i=1}^n  I(M_i > m)\widehat{W}_{i0}(t).
\end{align}

\begin{align*}
\mathrm{TPAUC}_t 
&= P\!\left(
M_{1} > M_{2}, \;
M_{1} \leq \mathrm{TPR}_t^{-1}(l_0), \;
M_{2} \geq \mathrm{FPR}_t^{-1}(u_0)
\;\middle|\;
T_{1} \leq t, \; T_{2} > t\right) \nonumber \\[1ex]
&= P\big(M_i > M_j,\, M_i \leq m_p,\, M_j \geq m_w \,\big|\, T_i \leq t,\, T_j > t \big) \nonumber\\[1ex]
&= \frac{
\mathbb{E}\Big[ I(M_i > M_j,\, M_i \leq m_p,\, M_j \geq m_w)\, I(T_i \leq t,\, T_j > t) \Big]
}{
\mathbb{E}\Big[ I(T_i \leq t,\, T_j > t) \Big]
}\end{align*}
\[
= \frac{
\mathbb{E}\Big[ I(M_i > M_j)\, I(M_i \leq m_p)\, I(M_j \geq m_w)\,
\mathbb{E}\big( I(T_i \leq t)\, I(T_j > t) \,\big|\, Y_i, \Delta_i, M_i, Y_j, \Delta_j, M_j \big) \Big]
}{
\mathbb{E}\Big[ \mathbb{E}\big( I(T_i \leq t)\, I(T_j > t) \,\big|\, Y_i, \Delta_i, M_i, Y_j, \Delta_j, M_j \big) \Big]
}
\]

\begin{align*}    
=  \frac{
\sum_i \sum_j I(M_i > M_j)\, I(M_i \leq m_p)\, I(M_j \geq m_w)\,
 {W}_{i1}(t){W}_{j0}(t)
}{
\sum_i \sum_j {W}_{i1}(t){W}_{j0}(t)
} 
\quad [\text{Replacing the expectations}\\ \text{ by empirical average}]
\end{align*}
Replacing the weights $W_{i1}$, $W_{j0}$ and $m_p$, $m_w$ by their corresponding estimates we obtain,
\begin{align}
   \widehat{\mathrm{TPAUC}}_{t} = \frac{ \sum_{i=1}^n \sum_{j=1}^n \widehat{W}_{i1}(t)\widehat{W}_{j0}(t) 
\left[ I(M_i > M_j)  \right] 
I(M_i \leq \widehat m_p) I(M_j \geq \widehat m_w) }
{ \widehat{N}_0(t)\widehat{N}_1(t) }
\end{align}

\noindent using the above estimators of $\mathrm{TPR}_t(m)$ and $\mathrm{FPR}_t(m)$, $\mathrm{PJ_t}$ can be estimated.

\end{document}